\documentclass[superscriptaddress,twocolumn,floatfix,prb]{revtex4}

\usepackage{amsmath}
\usepackage{graphicx}
\usepackage[caption=false]{subfig}
\usepackage{commath}

\newcommand{\av}[1]{\langle #1 \rangle}
\newcommand{\nnsum}[2]{\sum_{\langle #1,\,#2 \rangle}}
\newcommand{\sav}[1]{\left[ #1 \right]_{\mathrm{av}}}
\newcommand{\signame}{range parameter}

\begin{document}

\title{Low-temperature behavior of the statistics of the overlap
distribution in Ising spin-glass models}

\author{Matthew Wittmann}
\affiliation{Department of Physics, University of California, Santa Cruz,
California 95064, USA}

\author{B. Yucesoy}
\affiliation{Physics Department, University of Massachusetts, Amherst,
Massachusetts 01003, USA}
\affiliation{Center for Complex Network Research,
Northeastern University, 360 Huntington Avenue, Boston, Massachusetts 02115, 
USA}

\author{Helmut G. Katzgraber}
\affiliation{Department of Physics and Astronomy, Texas A\&M University,
College Station, Texas 77843-4242, USA}
\affiliation{Materials Science and Engineering Program, Texas A\&M University,
College Station, Texas 77843, USA}
\affiliation{Santa Fe Institute, 1399 Hyde Park Road, Santa Fe, New Mexico
87501, USA}

\author{J. Machta}
\affiliation{Physics Department, University of Massachusetts, Amherst,
Massachusetts 01003, USA}
\affiliation{Santa Fe Institute, 1399 Hyde Park Road, Santa Fe, New Mexico 
87501, USA}

\author{A.~P.~Young}
\affiliation{Department of Physics, University of California, Santa Cruz,
California 95064, USA}

\begin{abstract}

Using Monte Carlo simulations, we study in detail the overlap
distribution for individual samples for several spin-glass models
including the infinite-range Sherrington-Kirkpatrick model, short-range
Edwards-Anderson models in three and four space dimensions, and
one-dimensional long-range models with diluted power-law interactions.
We study three long-range models with different powers as follows: The
first is approximately equivalent to a short-range model in three
dimensions, the second to a short-range model in four dimensions, and
the third to a short-range model in the mean-field regime. We study an
observable proposed earlier by some of us which aims to distinguish the
``replica symmetry breaking'' picture of the spin-glass phase from the
``droplet picture,'' finding that larger system sizes would be needed to
unambiguously determine which of these pictures describes the
low-temperature state of spin glasses best, except for the
Sherrington-Kirkpatrick model, which is unambiguously described by
replica symmetry breaking.  Finally, we also study the median integrated
overlap probability distribution and a typical overlap distribution,
finding that these observables are not particularly helpful in
distinguishing the replica symmetry breaking and the droplet pictures.

\end{abstract}


\maketitle

\section{Introduction}

Despite much debate, there is still no consensus as to the nature of the
spin-glass state. According to the ``replica symmetry breaking'' (RSB)
picture of Parisi,\cite{parisi:79,parisi:80,parisi:83} there are many
``pure states,'' a nontrivial order parameter distribution, and a line
of transitions in a magnetic field, the de Almeida-Thouless
(AT)\cite{almeida:78} line.  By contrast, according to the droplet
theory,\cite{fisher:86,fisher:87,fisher:88,bray:86,mcmillan:84a} there
is only a symmetry-related pair of pure states in zero field (one state
in a nonzero field), the order parameter distribution is trivial in the
thermodynamic limit, and there is no AT line. The nature of the
spin-glass state has been investigated in a series of papers by Newman
and Stein (see, for example, Ref.~\onlinecite{newman:07} and references
therein), and most recently in a paper by Read.\cite{read:14a} A
discussion from an RSB point of view can be found in
Ref.~\onlinecite{marinari:00a}.

The averaged order parameter distribution $P(q)$, defined in
Eqs.~\eqref{PJq} and \eqref{Pavq} below, is predicted to be nonzero in
the vicinity of $q=0$ as the size of the system $N \equiv L^d$ tends to
infinity according to RSB theory,\cite{parisi:83} whereas it is expected
to vanish\cite{fisher:86} as $L^{-\theta}$ in the droplet picture where
$\theta$ is a positive ``stiffness'' exponent. Results from
simulations\cite{reger:90,marinari:00a,katzgraber:01,katzgraber:03} seem
close to the predictions of RSB, but it has been
argued\cite{moore:98,middleton:13} that the sizes which can be simulated
are too small to see the asymptotic behavior.

Consequently, there has recently been
interest\cite{yucesoy:12,middleton:13,monthus:13} in studying other
quantities related to $P(q)$ but where more attention is paid to the
overlap distribution of \textit{individual} samples, $P_{\mathcal
J}(q)$, rather than just calculating the sample \textit{average}.
Accurately determining $P_{\mathcal J}(q)$ for each sample is more
demanding numerically than just computing the average, but computer
power has advanced to the point where this is now feasible.

In this paper we study in detail these new quantities for a
\textit{range} of models. In addition to short-range Edwards-Anderson
(EA) Ising spin-glass models in three (3D) and four (4D) space
dimensions, and the infinite-range
Sherrington-Kirkpatrick\cite{sherrington:75} (SK) model, we also study
diluted long-range (LR) Ising spin-glass models in one space dimension
(1D) in which the interaction falls off with a power of the distance
between two spins.  Varying the power is argued to be analogous to
changing the space dimension $d$ of a short-range
model.\cite{katzgraber:03,katzgraber:05cKY,katzgraber:08,larson:10,leuzzi:08,banos:12}
An important advantage of the LR models is that one can study them in
effective space dimensions $d \geq 6$ that are not easily accessible for
short-range models via computer simulations.  In this regime the number
of spins for short-range (SR) models $N = L^d$ increases so fast with
the linear system size $L$ that one cannot simulate the range of sizes
that is necessary for finite-size scaling (FSS). It is important to
study $d \geq 6$ because it is conjectured that $d=6$ is the upper
critical dimension above which mean-field behavior is seen. Finally,
verifying the consistency of our results for both SR and LR models gives
us additional confidence in our numerical results.

The plan of this paper is as follows. Section \ref{sec:models} describes
the several models that we study, while Sec.~\ref{sec:methods} discusses
the Monte Carlo technique. In Sec.~\ref{sec:qoi} we explain the
quantities we compute to try to understand better the nature of the
spin-glass state, and the results are given in Sec.~\ref{sec:results}.
We summarize our results and give our conclusions in
Sec.~\ref{sec:concl}.

\section{Models}
\label{sec:models}

We study several classes of Ising spin-glass models.  These are
long-range one-dimensional models,  three- and four-dimensional
short-range models known as  Edwards-Anderson models, and the
infinite-range spin glass known as the Sherrington-Kirkpatrick
model.  In all cases the Hamiltonian can be written in the form
\begin{equation}
\mathcal{H} = - \sum_{i,\, j} J_{i j} S_i S_j\, ,
\label{ham}
\end{equation}
where the $S_i$ ($i = 1,\, 2,\, \dots,\, N$) represent Ising spins that
take values $\pm 1$, and the $J_{i j}$ are statistically independent,
quenched random variables. The summation is defined over all pairs of
interacting spins. All of the models studied here have
finite-temperature spin-glass transitions.  The models differ according
to which spins interact and the strength of the couplings.

\subsection{Edwards-Anderson models on hypercubic lattices}

The three- and four-dimensional EA models that we study are defined on
(hyper)cubic lattices with periodic boundary conditions. The
nearest-neighbor interactions are chosen from a Gaussian distribution
with zero mean and unit variance,
\begin{equation}
\sav{J_{i j}}=0 \;\;\;\;\;\;\;\;\;\; \sav{J_{i j}^2}=1 \, ,
\end{equation}
where $\sav{\ldots}$ indicates a quenched average over the couplings.
From numerical studies it is known that the transition temperatures are
$T_c = 0.951(9)$\cite{katzgraber:06} in 3D and $T_c =
1.80(1)$\cite{parisi:96} in 4D.

\subsection{Sherrington-Kirkpatrick model}

For the SK\cite{sherrington:75} model each
spin interacts with every other spin. A coupling is chosen from a
Gaussian distribution with zero mean and variance,
\begin{equation}
\sav{J_{i j}^2}=1/N .
\end{equation}
The variance of the coupling is inversely proportional to the number of
spins $N$ so that there is a well-defined thermodynamic limit. The
transition temperature for this model is $T_c = 1$.\cite{sherrington:75}

\subsection{One-dimensional diluted long-range  model}

For the diluted LR models the mean coupling is zero but the variance
depends on the distance between the spins according to
\begin{equation}
\sav{J_{i j}^2} \propto r_{i j}^{-2 \sigma}\, ,
\label{var}
\end{equation}
where $\sigma$ is the \signame, and $r_{i j}$ is the chord distance
between sites $i$ and $j$ when the sites are arranged on a
ring,\cite{katzgraber:03} i.e.,
\begin{equation}
r_{i j} = \frac{N}{\pi} \sin \left( \frac{ \pi \abs{i - j}}{N} \right).
\end{equation}
We choose a distribution $P(J_{i j})$ that satisfies Eq.~\eqref{var} at a
large distance while allowing for efficient computer simulation. In this
\emph{diluted model},\cite{leuzzi:08,katzgraber:09b} most of the
interactions between two spins are absent (i.e., most of the $J_{i j}$
are zero) and it is the \emph{probability} of there being a bond between
two spins (rather than its strength) that falls off with their
separation (asymptotically as $1/r_{i j}^{2 \sigma}$). More precisely,
\begin{equation}
P(J_{i j}) = (1 - p_{i j}) \delta(J_{i j})
+ p_{i j} \frac{1}{\sqrt{2 \pi}} e^{-J_{i j}^2 / 2},
\label{P_dil}
\end{equation}
where $p_{i j} \propto 1/r_{i j}^{2\sigma}$ at a large distance.  It is
convenient to fix the mean number of neighbors $z$. The pairs of sites
with nonzero bonds are then generated as follows. Pick a site $i$ at
random.  Then pick a site $j$ with probability $\widetilde{p}_{i j} =
A/r_{i j}^{2 \sigma}$, where $A$ is determined by
normalization.\cite{katzgraber:09b} If there is already a bond between
$i$ and $j$, repeat until a pair $i$, $j$ is selected which does not
already have a bond.\cite{comment:norm} At that point set $J_{i j}$
equal to a Gaussian random variable with mean zero and variance unity.
This process is repeated $N z/2$ times so the number of sites connected
to a given site has a Poisson distribution with mean $z$.  Because each
site has, on average, $z$ neighbors, and the variance of each
interaction is unity, we have
\begin{equation}
\sum_j \sav{J_{i j}^2} = z.
\label{constraint_dil}
\end{equation}
This prescription has the advantage that Monte Carlo updates  require
only a time proportional to $N z$ rather than $N^2$ that would be
required if all bonds were present.\cite{leuzzi:08,katzgraber:09b}

We consider three values of the \signame: $\sigma=0.6$, which is in the
mean-field region,\cite{larson:10} $\sigma=0.784$, which represents, at
least approximately, a short-range system in four
dimensions,\cite{katzgraber:09,larson:10,larson:13,banos:12} and
$\sigma=0.896$ which approximately represents a three-dimensional
system.\cite{katzgraber:09,larson:10,larson:13,banos:12} The values of
$T_c$ are approximately equal to\cite{larson:13} $1.35$ and $0.795$ for
$\sigma = 0.784$ and $0.896$, respectively.  For $\sigma = 0.6$ we find
$T_c \approx 1.953$.

\section{Methods}
\label{sec:methods}

We have carried out parallel tempering/replica-exchange Monte Carlo
simulations\cite{geyer:91,hukushima:96,marinari:98b} of the models
described in Sec.~\ref{sec:models}.  In parallel tempering, $N_T$
replicas of the system with the same couplings are each simulated at a
different temperature in the range $T_{\rm min}$ -- $T_{\rm max}$.  In
addition to standard Metropolis sweeps at each temperature, there are
parallel tempering moves that allow replicas to be exchanged between
neighboring temperatures.  A single sweep consists of a Metropolis sweep
at each temperature, followed by a set of parallel tempering moves between
each pair of neighboring temperatures.  The power of parallel tempering
is that the temperature swap moves permit replicas to diffuse from low
temperatures, where equilibration is very difficult, to high temperatures,
where it is easy, and back to low temperature.  These round trips greatly
accelerate equilibration at the lowest temperatures.  The simulation
parameters are shown in Tables~\ref{tab:params}--\ref{tab:paramssk}.
The parameter $b$ determines the number of sweeps: $2^b$ for
equilibration followed by $2^b$ for data collection.  The parameter
$N_{\rm sa}$ is the number of disorder samples simulated.

For each model we have chosen the lowest temperature to be less than or
equal to $0.4 T_c$, the approximate temperature for which we report most
of our results.

\begin{table}[htb]
\caption{
Simulation parameters for the 1D models. For each value of $\sigma$ and
size $N$, $N_{\mathrm{sa}}$ samples were equilibrated for $2^b$ sweeps and
then measured for an addition $2^b$ sweeps,
using replica-exchange Monte Carlo with $N_T$ temperatures distributed
between $T_{\mathrm{min}}$ and $T_{\mathrm{max}}$.
}
   \begin{ruledtabular}
    \begin{tabular}{rrrrrrr}
      $\sigma$ & $N$ & $b$ & $T_{\mathrm{min}}$ &
      $T_{\mathrm{max}}$ & $N_T$ & $N_{\mathrm{sa}}$ \\
      \hline
      0.6 & 64 & 24 & 0.82 & 3 & 50 & 4992 \\
      0.6 & 128 & 24 & 0.82 & 3 & 50 & 4800 \\
      0.6 & 256 & 24 & 0.82 & 3 & 50 & 4800 \\
      0.6 & 512 & 24 & 0.82 & 3 & 50 & 4684 \\
      0.6 & 1024 & 25 & 0.82 & 3 & 50 & 4800 \\[2mm] 
      0.784 & 64 & 24 & 0.55 & 2 & 50 & 4377 \\
      0.784 & 128 & 24 & 0.55 & 2 & 50 & 5060 \\
      0.784 & 256 & 24 & 0.55 & 2 & 50 & 5470 \\
      0.784 & 512 & 24 & 0.55 & 2 & 50 & 5207 \\
      0.784 & 1024 & 25 & 0.55 & 2 & 50 & 5988 \\[2mm] 
      0.896 & 64 & 24 & 0.31 & 1.2 & 50 & 2600 \\
      0.896 & 128 & 24 & 0.31 & 1.2 & 50 & 4468 \\
      0.896 & 256 & 24 & 0.31 & 1.2 & 50 & 4749 \\
      0.896 & 512 & 24 & 0.31 & 1.2 & 50 & 4749 \\
      0.896 & 1024 & 25 & 0.31 & 1.1788 & 25 & 4749 \\
    \end{tabular}
   \end{ruledtabular}
    \label{tab:params}
\end{table}

\begin{table}[htb]
\caption{
Simulation parameters for the 3D EA spin glass. For each number of spins
$N= L^3$ we equilibrate and measure for $2^b$ Monte Carlo sweeps.
$T_{\rm min}$ [$T_{\rm max}$] is the lowest [highest] temperature used
and $N_T$ is the number of temperatures. $N_{\rm sa}$ is the number of
disorder samples. For $T \ge 0.42$ all system sizes are in thermal
equilibrium.
\label{tab:paramsea3}}
\begin{tabular*}{\columnwidth}{@{\extracolsep{\fill}} r r c c c c c}
\hline
\hline
$N$ &  $L$ & $b$  & $T_{\rm min}$ & $T_{\rm max}$ & $N_{T}$ & $N_{\rm sa}$ \\
\hline
  $64$ &  $4$ & $18$ & $0.2000$       & $2.0000$        & $16$    & $4891$ \\
 $216$ &  $6$ & $24$ & $0.2000$       & $2.0000$        & $16$    & $4961$ \\
 $512$ &  $8$ & $27$ & $0.2000$       & $2.0000$        & $16$    & $5130$ \\
$1000$ & $10$ & $27$ & $0.2000$       & $2.0000$        & $16$    & $5027$ \\
$1728$ & $12$ & $25$ & $0.4200$       & $1.8000$        & $26$    & $3257$ \\
\hline
\hline
\end{tabular*}

\caption{
Simulation parameters for the 4D EA spin glass. For details, see the
caption of Table \ref{tab:paramsea3}.  Here $N=L^4$.
\label{tab:paramsea4}}
\begin{tabular*}{\columnwidth}{@{\extracolsep{\fill}} r r c c c c c}
\hline
\hline
$N$ &  $L$ & $b$  & $T_{\rm min}$ & $T_{\rm max}$ & $N_{T}$ & $N_{\rm sa}$ \\
\hline
  $256$ &  $4$ & $23$ & $0.7200$       & $2.3800$        & $52$    & $3252$ \\
 $625$ &  $5$ & $23$ & $0.9101$       & $2.3800$        & $42$    & $4086$ \\
 $1296$ &  $6$ & $23$ & $0.7200$       & $2.3800$        & $52$    & $3282$ \\
$2401$ & $7$ & $23$ & $0.9101$       & $2.3800$        & $42$    & $4274$ \\
$4096$ & $8$ & $23$ & $0.7200$       & $2.3800$        & $52$    & $3074$ \\
$6561$ & $9$ & $24$ & $0.7200$       & $2.3800$        & $52$    & $3010$ \\
\hline
\hline
\end{tabular*}

\caption{
Simulation parameters for the SK spin glass. For details, see the
caption of Table \ref{tab:paramsea3}.
\label{tab:paramssk}}
\begin{tabular*}{\columnwidth}{@{\extracolsep{\fill}} r c c c c c}
\hline
\hline
   $N$ & $b$  & $T_{\rm min}$  & $T_{\rm max}$   & $N_{T}$ & $N_{\rm sa}$ \\
\hline
  $64$ & $22$ & $0.2000$       & $1.5000$        & $48$    & $5068$       \\
 $128$ & $22$ & $0.2000$       & $1.5000$        & $48$    & $5302$       \\
 $256$ & $22$ & $0.2000$       & $1.5000$        & $48$    & $5085$       \\
 $512$ & $18$ & $0.2000$       & $1.5000$        & $48$    & $4989$       \\
$1024$ & $18$ & $0.2000$       & $1.5000$        & $48$    & $3054$       \\
$2048$ & $16$ & $0.4231$       & $1.5000$        & $34$    & $3020$       \\
\hline
\hline
\end{tabular*}
\end{table}

To test our simulations for equilibration, we use an \emph{equilibrium}
relationship between sample-averaged quantities, valid for systems with
Gaussian interactions, which has been discussed
before.\cite{katzgraber:01,katzgraber:03} Except for the SK model, the relation
is:\cite{katzgraber:03}
\begin{equation}
U = -\frac{z}{2T}(1-q_l),
\label{equil}
\end{equation}
where $z$ is the (mean) coordination number, $T$ is the temperature, 
\begin{equation}
U = -\frac{1}{N}\nnsum{i}{j} \sav{J_{i j}\langle S_i S_j \rangle}
\end{equation}
is the energy per spin and $q_l$ is the \emph{link overlap},
\begin{equation}
q_l = \frac{2}{Nz}\nnsum{i}{j}\sav{\epsilon_{i j} \av{S_i S_j}^2},
\label{link-overlap}
\end{equation}
where $\epsilon_{i j}=1$ if there is a bond between $i$ and $j$ and zero
otherwise.  For the SK model one obtains formally the corresponding
relation by putting $z=1$ in Eqs.~\eqref{equil} and \eqref{link-overlap}
and setting all the $\epsilon_{ij}$ equal to $1$.  For the EA models $z=2d$
and $\epsilon_{ij}$ is defined by the associated hypercubic lattices
with periodic boundary conditions.

While Eq.~\eqref{equil} is a useful criterion for the equilibration of
sample-averaged quantities such as the energy and overlap, we must be
more careful when studying quantities that may be sensitive to the
equilibration of individual samples, such as those considered in
Sec.~\ref{sec:qoi}. To study such quantities we run our simulations for
many times the number of sweeps needed to satisfy Eq.~\eqref{equil}.  In
fact, we require that at least three logarithmically spaced bins agree
within error bars.

\begin{figure}
\centering
\includegraphics{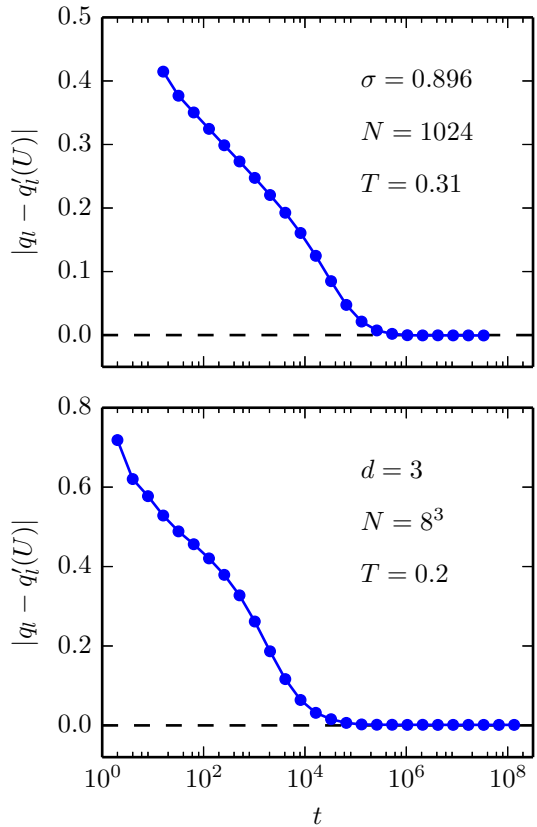}
\caption{(Color online)
Absolute difference between $q_l$ and the quantity $q_l^{\prime}(U)
\equiv (2T/z)U + 1$, obtained from the equilibrium relationship of
Eq.~\eqref{equil}, for the LR model with $N=1024, \sigma=0.896$ (top
panel) and the 3D EA model with $L=8$ (lower panel) as a function of
Monte Carlo sweeps ($t$) on a log-linear scale. At large times the
difference is zero, but the simulation continues well beyond this point
to ensure that good statistics are obtained for all samples.
Error bars are smaller than the symbols. Both panels have the same
horizontal scale.
}\label{fig:equilibration}
\end{figure}

Figure \ref{fig:equilibration} shows an example of the equilibration
test for the 1D LR model with $\sigma=0.896$ for the largest size at the
lowest temperature, and also for the 3D EA model with $L=8$, again at
the lowest temperature.  The vertical axis is the difference between the
two sides of Eq.~\eqref{equil} while the horizontal axis is the number
of Monte Carlo sweeps on a logarithmic scale. This difference vanishes
within the error bars at around $10^5$ sweeps in both cases but the
simulation continues for much longer than this to ensure that good
statistics are obtained for \textit{all} samples.

\begin{figure*}
  \subfloat[$\Delta(q_0=0.2,\,\kappa=1)$]{
    \includegraphics{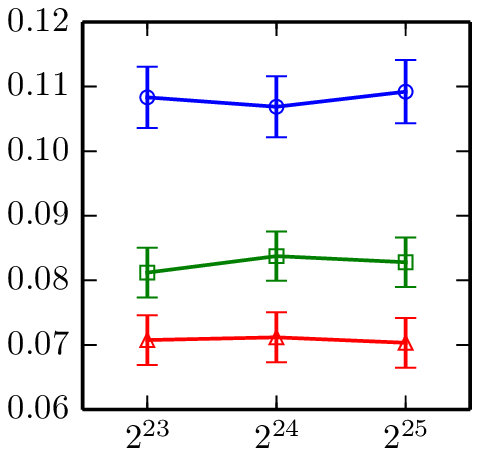}
  }
  \subfloat[$I^\text{av}(q=0.2)$]{
    \includegraphics{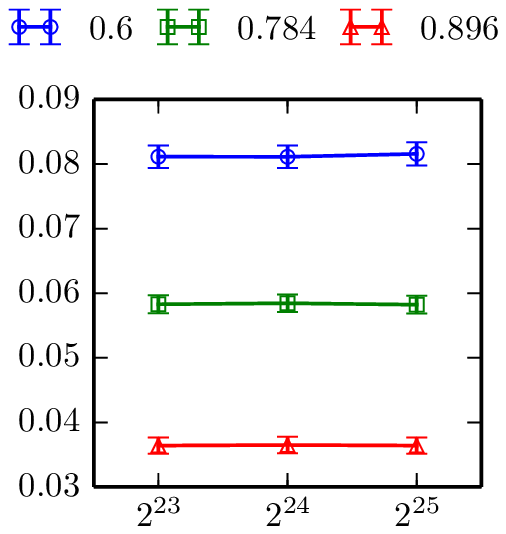}
  }
  \subfloat[$I^\text{med}(q=0.2)$]{
    \includegraphics{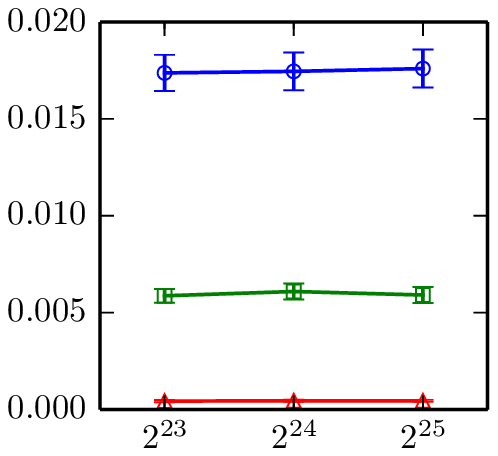}
  }
\caption{(Color online)
Plots of several observables obtained from the overlap distribution,
defined in Sec.~\ref{sec:qoi}, vs the number of Monte Carlo sweeps for
the largest size studied, $N=1024$, for the long-range model at the
lowest temperature simulated for each value of $\sigma$.  (See
Table~\ref{tab:params}.)
}
  \label{fig:equil-qoi}
\end{figure*}

As an additional check on equilibration for the 1D LR models, Fig.
\ref{fig:equil-qoi} shows  several quantities of interest, defined in
Sec.~\ref{sec:qoi}, as a function of the number of sweeps on a log
scale, for the lowest temperature studied and for each value of
$\sigma$. The data appear to have saturated.

The 3D EA data set has also been tested for equilibration using the
integrated autocorrelation time, as discussed in
Ref.~\onlinecite{yucesoy:13}.

\section{Measured Quantities}
\label{sec:qoi}

For a single sample
${\mathcal J}\equiv \{J_{ij}\}$,
the spin overlap
distribution $P_{\mathcal J}(q)$ is given by
\begin{equation}
P_{\mathcal J}(q) = \left\langle \delta \left(\,q - 
{1\over N} \sum_{i=1}^N S_i^{(1)} S_i^{(2)}\,
\right) \right\rangle
\, ,
\label{PJq}
\end{equation}
where ``$(1)$'' and ``$(2)$'' refer to two independent copies of the
system with the same interactions, and $\langle \cdots
\rangle$ denotes a thermal (i.e., Monte Carlo) average for the single sample.
In most previous work, $P_{\mathcal J}(q)$ is simply
averaged over disorder samples to obtain $P(q)$ defined by
\begin{equation}
P(q) = \sav{P_{\mathcal J}(q)} \, .
\label{Pavq}
\end{equation}

In order to gain additional information that might distinguish the RSB
and droplet pictures, several investigators have recently introduced
other observables related to the statistics of $P_{\mathcal J}(q)$.
Yucesoy {\em et al}.\cite{yucesoy:12} proposed a  measure that is
sensitive to peaks in the overlap distributions of \textit{individual
samples}. A sample is counted as ``peaked'' if $P_{\mathcal{J}}(q)$
exceeds a threshold value $\kappa$ in the domain $\abs{q}<q_0$. The
quantity $\Delta(q_0,\,\kappa)$ is then defined as the fraction of
peaked samples. More precisely, for each sample let
\begin{equation}
  \Delta_{\mathcal{J}}(q_0,\,\kappa) =
  \begin{cases}
    1, &  \mathrm{if}\  P_{\mathcal{J}}^{\mathrm{max}}(q_0) > \kappa , \\
    0, & \mathrm{otherwise,}
  \end{cases}
  \label{eq:deltaj}
\end{equation}
where $P_{\mathcal{J}}^{\mathrm{max}}(q_0)$ is the maximum value of the
distribution in the domain specified by $q_0$,
\begin{equation}
P_{\mathcal{J}}^{\mathrm{max}}(q_0) = 
\max \{ P_{\mathcal{J}}(q):\,\abs{q}<q_0 \}.
\label{eq:pjmax}
\end{equation}
We then define $\Delta(q_0,\,\kappa)$ to be the sample average,
\begin{equation}
  \Delta(q_0,\,\kappa) = \sav{\Delta_{\mathcal{J}}(q_0,\,\kappa)}.
  \label{eq:delta}
\end{equation}
The quantity $\Delta(q_0,\kappa)$ is a nondecreasing function of $q_0$
and a nonincreasing function of $\kappa$.  This behavior follows simply
from the definition of $\Delta(q_0,\kappa)$.  A more important property
of $\Delta(q_0,\,\kappa)$ is that it must go either to zero or one as $N
\rightarrow \infty$.\cite{comment:newman+stein} All the scenarios for
the low-temperature behavior of spin-glass models predict that
$P_{\mathcal{J}}(q)$ consists of $\delta$ functions as $N \rightarrow
\infty$.  The difference between scenarios lies in the number and
position of these $\delta$ functions.  The RSB picture predicts that
there is a countable infinity of $\delta$ functions that densely fill
the line between $-q_{\rm EA}$ and $+q_{\rm EA}$.  Thus, for any $q_0$
and any $\kappa$, $\Delta(q_0,\,\kappa) \rightarrow 1$ for models
described by RSB.  On the other hand, for models described by the
droplet scenario or other single pair of states scenarios,
$\Delta(q_0,\,\kappa) \rightarrow 0$ for any $q_0 < q_{\rm EA}$ and any
$\kappa$.  Thus, the quantity $\Delta(q_0,\,\kappa)$ will sharply
distinguish the RSB and droplet scenarios if one can study large enough
sizes.  We shall study the size dependence of $\Delta$ numerically for
all our models in Sec.~\ref{sec:Delta}.

As mentioned above, most previous work evaluated the \textit{average}
probability distribution $P(q)$, but recently
Middleton,\cite{middleton:13} and Monthus and Garel,\cite{monthus:13}
have proposed measures yielding a \textit{typical} value of the sample
distribution $P_{\mathcal J}(q)$ in the hopes that these measures would
provide a clearer differentiation between the RSB and droplet pictures
than the average $P(q)$.

Middleton\cite{middleton:13} studied  $I^\text{med}(q)$, the
\textit{median} of the \textit{cumulative} overlap distribution of a
single sample $I_{\mathcal{J}}(q)$, where $I_{\mathcal{J}}(q)$ is
defined by
\begin{equation}
I_{\mathcal{J}}(q) = \int_{-q}^{q} P_{\mathcal{J}}(q') \dif q'.
\end{equation}
We also denote the average cumulative distribution over samples by 
$I^\text{av}(q)$, which is given by
\begin{equation}
I^\text{av}(q) = \int_{-q}^{q} P(q') \dif q' \, .
\label{Iav}
\end{equation}
The median is insensitive to the effect of samples with unusually large
values of $I_{\mathcal{J}}(q)$.

For the SK model $P(q)$ tends to a constant as $q \to 0$, and so
$I^\text{av}(q) \propto q$ for small $q$.  We can obtain a rough idea of
how $I^\text{med}(q)$ varies with $q$ for small $q$ in the SK model from
the results of M\'ezard {\em et al}.\cite{mezard:84} First of all, to
obtain a notation which is more compact and is extensively used in other
work, we write $x(q) \equiv I^\text{av}(q)$.  M\'ezard {\em et
al}.\cite{mezard:84} argue that, at small $q$ where $x(q)$ is also
small, the probability of a certain integrated value $I_{\mathcal{J}}$
is given by
\begin{equation}
p(I_{\mathcal{J}}) \propto x\, I_{\mathcal{J}}^{x-1}\, ,
\label{pIJ}
\end{equation}
where we recall that $x$ is the average value of $I_{\mathcal{J}}$.
From Eq.~\eqref{pIJ} we estimate the median in terms of the average as
\begin{equation}
I^\text{med}(q) \propto e^{-\ln 2/x(q)} = e^{-\ln 2/[2 q P(0)]}
\label{sk-median-theory}
\end{equation}
for $q \to 0$, where we used that $P(0)$ is nonzero so $x(q) \equiv
I^\text{av}(q) \simeq 2 P(0)\, q$ in this limit [see Eq.~\eqref{Iav}].
Hence the median tends to zero exponentially fast as $q \to 0$ whereas
the average only goes to zero linearly.

\begin{figure*}
\includegraphics{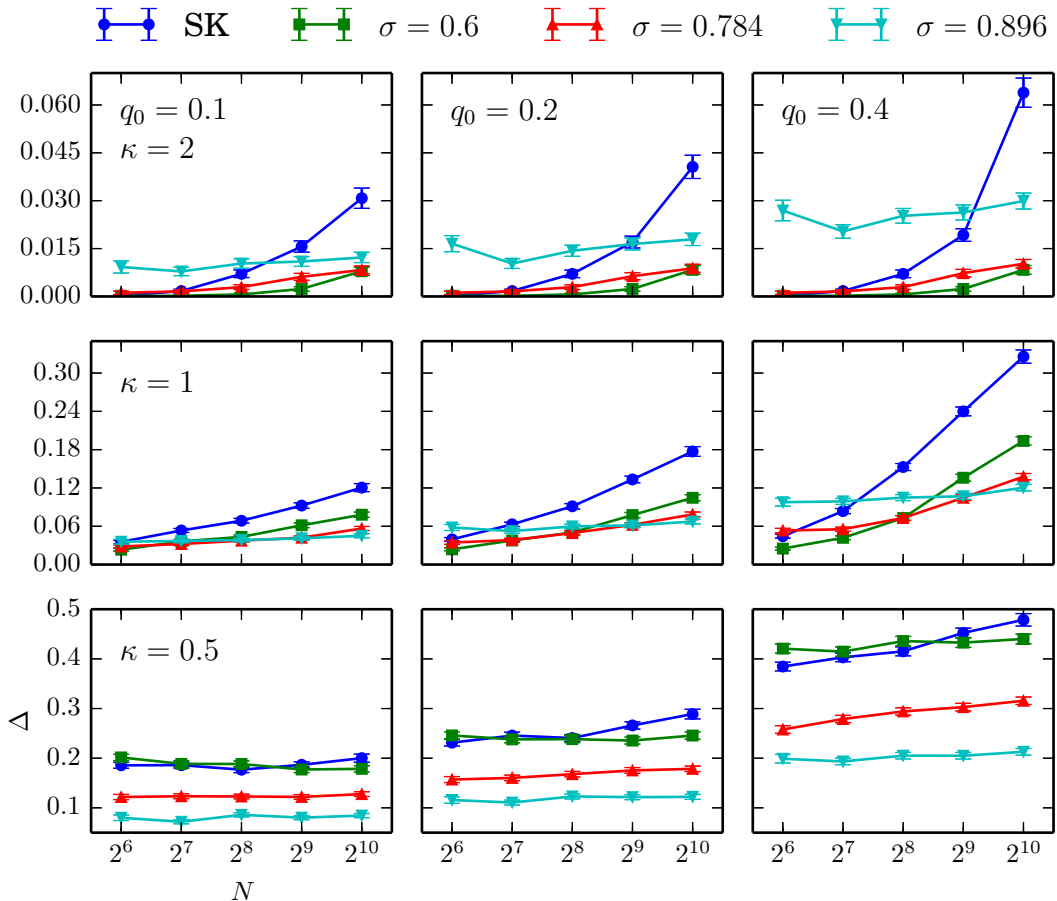}
\caption{(Color online)
$\Delta(q_0,\,\kappa)$ as a function of system size $N$ for the
long-range models and the SK model for all available values of $\sigma$
and various values of the window $q_0$ and threshold $\kappa$. In all
cases the temperature is $0.4 T_c$. 
All panels have the same
horizontal scale, and all panels in a row have the same vertical scale.
}
\label{fig:delta-vs-N-grid}
\end{figure*}

In the droplet picture, $P(0)$ is expected to vanish with $L$
as\cite{fisher:86} $L^{-\theta}$, so $I^\text{av}(q) \propto
L^{-\theta}\, q$ for small $q$.  The median value $I^\text{med}(q)$ will
presumably also vanish for small $q$ as $L \to \infty$, but we are not
aware of any precise predictions for this.  We shall study the median
cumulative distribution numerically in Sec.~\ref{sec:median}.

Another measure related to the overlap distribution of individual
samples has been proposed by Monthus and Garel.\cite{monthus:13} They
suggest calculating a ``typical" overlap distribution defined by the
exponential of the average of the log as
\begin{equation}
\label{eq:typ}
P^\text{typ}(q) =  \exp\sav{\ln P_{\mathcal{J}}(q)} \, .
\end{equation}
We shall study this quantity numerically in Sec.~\ref{sec:typical}.

\begin{figure*}
\includegraphics{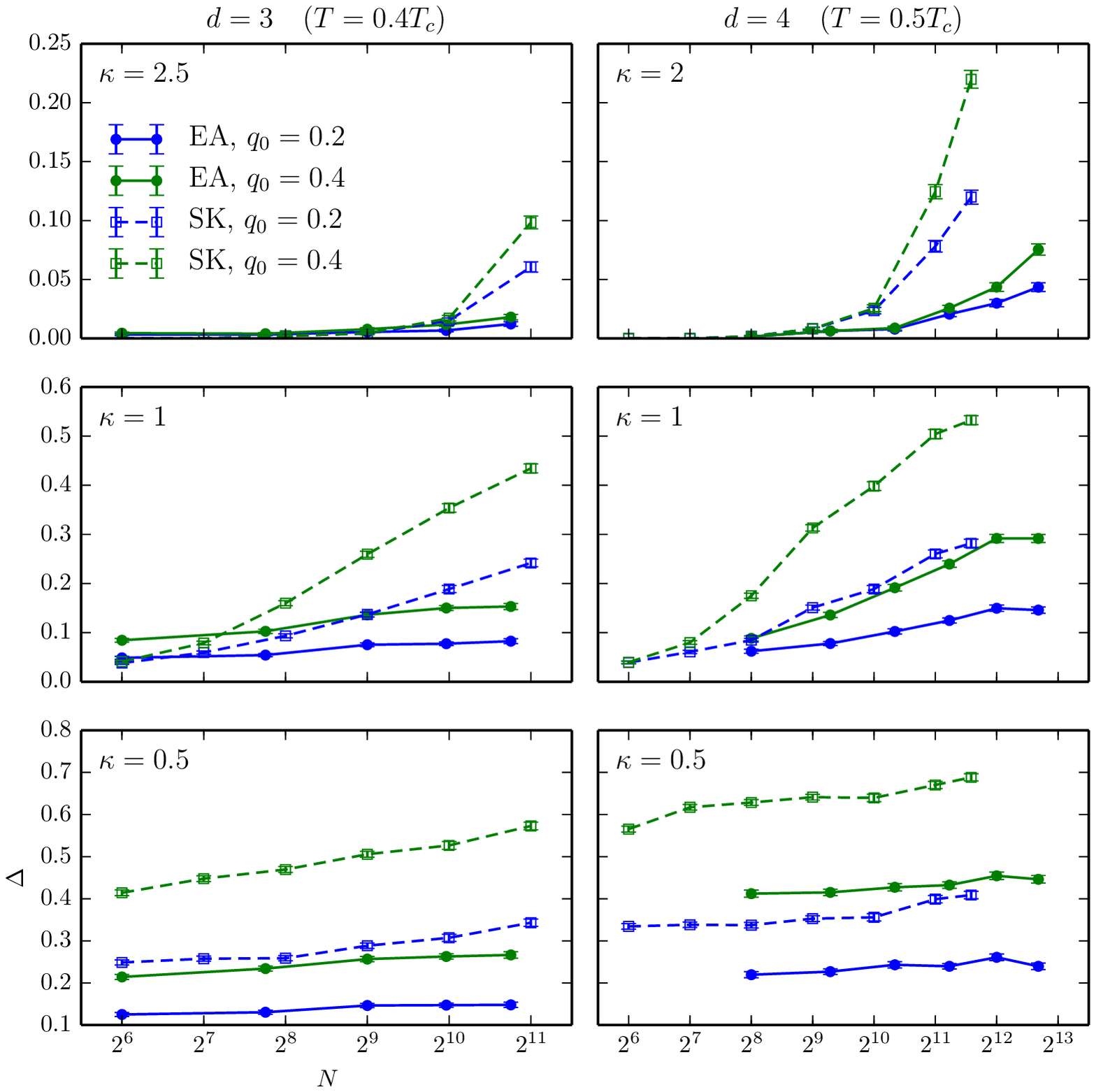}
\vspace*{-2.5em}
\caption{(Color online)
$\Delta(q_0,\,\kappa)$ as a function of system size $N$ for the
short-range models and the SK model for several values of the window
$q_0$ and threshold value $\kappa$. The points connected by solid lines
are for the EA models, while those connected by dashed lines are for the
SK model. The temperatures are $0.4 T_c$ for the 3D data and $0.5 T_c$
for the 4D data. 
All panels in a column have the same 
horizontal scale and all panels in a row have the same vertical scale.
}
\label{fig:delta-vs-N-grid-EA-short}
\end{figure*}

\section{Results}
\label{sec:results}

\subsection{Fraction of peaked samples, $\Delta(q_0,\kappa)$}
\label{sec:Delta}

Plots of $\Delta(q_0,\kappa)$ for the 1D long-range models for various
values of $q_0$ and $\kappa$ at $T \approx 0.4 T_c$ are given in
Fig.~\ref{fig:delta-vs-N-grid}, while the corresponding plots for the 3D
and 4D models are shown in Fig.~\ref{fig:delta-vs-N-grid-EA-short}. A
comparison with the SK model is made in both cases.  The error bars for
all plots in this section  are one standard deviation statistical errors
due to the finite number of samples.  There are also errors in the data
for each sample due to the finite length of the data collection. For the
EA and SK models, we estimated these errors by measuring
$\Delta^+(q_0,\kappa)$ and $\Delta^-(q_0,\kappa)$, defined as in
Eqs.~(\ref{eq:deltaj}) -- (\ref{eq:delta}) but from the $q>0$ and $q<0$
components of $P_{\mathcal{J}}(q)$, respectively. These are expected to
be reasonably independent and  their differences provide an estimate of
the error due to finite run lengths. For all sizes, the average absolute
difference between these quantities, $[|\Delta^+(q_0,\kappa) -
\Delta(q_0,\kappa)| + |\Delta^-(q_0,\kappa) - \Delta(q_0,\kappa)|]/2$, is
less than the statistical error. While a similar analysis was not done
for the 1D LR models, measurements of $\Delta$ versus the number of
sweeps shown in Fig.~\ref{fig:equil-qoi} suggest that the data have
saturated within statistical error.

One can draw several qualitative conclusions from these plots.   It is
apparent that $\Delta(q_0,\kappa)$ is an increasing function of $N$ for
small $N$.  As the system size increases, we expect $\Delta(q_0,\kappa)$
to increase because all the features of $P_{\mathcal{J}}(q)$ sharpen.
For the SK model, which is indisputably described by the  RSB picture,
the number of features and their height should both increase and
$\Delta(q_0,\kappa)$ should be a strongly increasing function of $N$.
Indeed, this behavior is seen except for $\kappa=0.5$, which is a
sufficiently small value that $\Delta(q_0,\kappa)$ is effectively
measuring whether or not there is a feature in the relevant range, and
this quantity increases relatively slowly for the SK model.

However, as $\sigma$ increases for the 1D models, the curves become
increasingly flat and the difference between $\sigma=0.896$ and the SK
model is striking; the former is nearly flat while the latter increases
sharply (see Fig.~\ref{fig:delta-vs-N-grid}). The same qualitative
distinction holds between the 3D EA model and the SK model (see
Fig.~\ref{fig:delta-vs-N-grid-EA-short}). The similarity between the
behavior of the 1D model for $\sigma=0.896$ and the 3D EA model is
expected since the two models are believed to have the same qualitative
behavior.  The distinction between the SK model and the 1D model with
$\sigma=0.784$ and the 4D EA model is less striking but qualitatively
the same.

It is interesting to compare results for the SK model with the 1D model
with $\sigma=0.6$, which is in the mean-field regime. For $\kappa = 0.5$
the results for the two models are very similar, and do not increase
much with $N$, indicating that $\kappa=0.5$ is too small to give useful
information for this range of sizes, as discussed above.  For $\kappa =
1$, the SK data increase the most rapidly with $N$, and the $\sigma =
0.6$ data increase less quickly, but still faster than the other values of
$\sigma$. For $\kappa = 2$, the SK data increase quickly, while for the
value of $\sigma$ furthest from the SK limit, $0.896$, the data are
moderately large but roughly size independent over the range of sizes
simulated.  Curiously, for intermediate values of $\sigma$ ($0.6$ and
$0.784$) the data are very small but show an increase for the larger
sizes. This increase is particularly sharp for $\sigma =0.6$.  It seems
that there is an initial value of $\Delta$ for small $N$ and a growth as
$N$ increases.  We do not have a good understanding of the initial
value, e.g., why it is so small for $\kappa = 2$ and $\sigma = 0.6$, and
$0.784$. The more important aspect of the data is the increase observed,
at least for most parameter values, at large sizes. Given the rapid
increase in the data for $\sigma = 0.6, \kappa=2$ for the largest size,
we anticipate that for still larger sizes, its value for $\Delta$ for
$\kappa = 2$ would be closer to that of the SK model than that of the
intermediate $\sigma$ values.

\begin{figure*}
\includegraphics{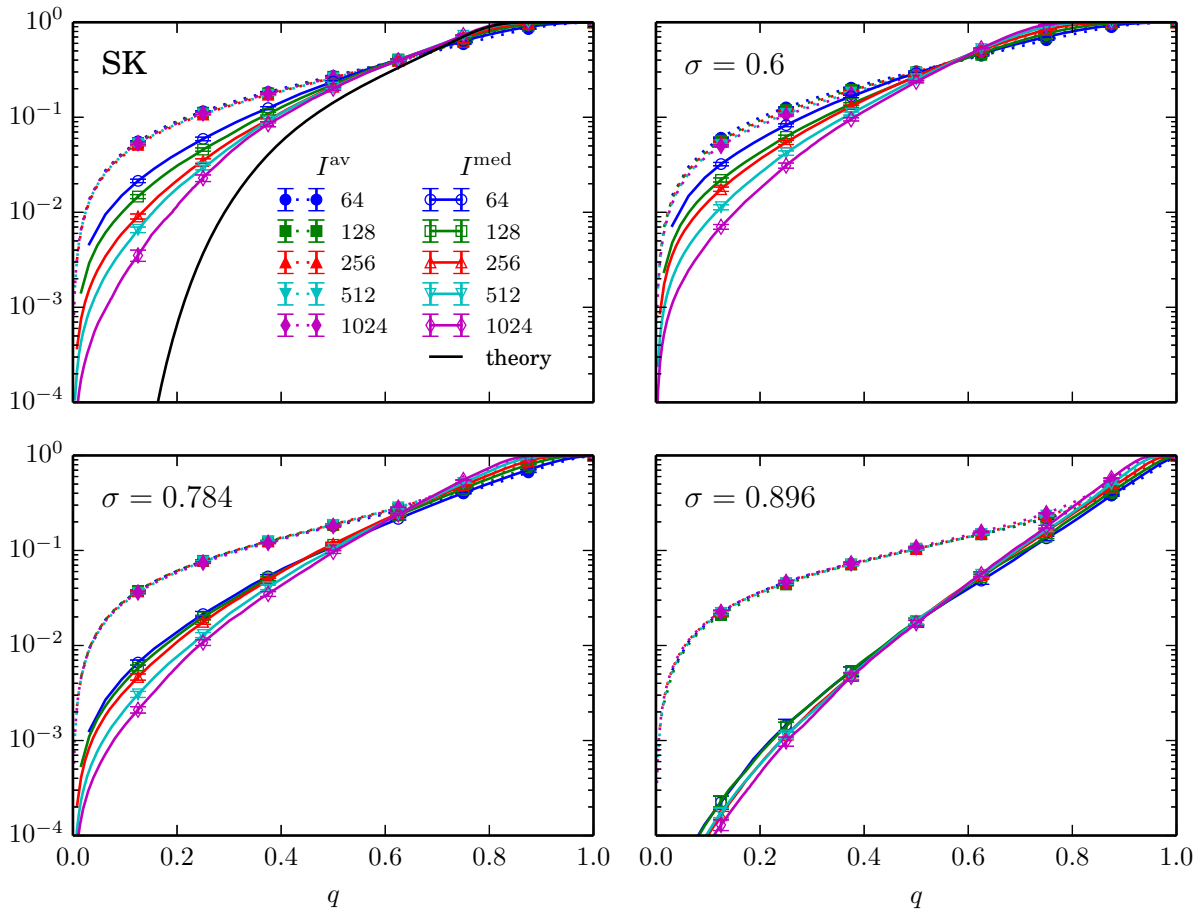}
\caption{(Color online)
Mean and median over samples of the integrated distribution
$I_{\mathcal{J}}(q)$ for the long-range models and the SK model.  In all
cases the temperature is close to $0.4 T_c$.  For both the SK model and
the 1D models, the median shows a relatively strong size dependence
compared with the mean, this difference being the least pronounced for
$\sigma = 0.896$. The ``theory'' curve for the SK data
[Eq.~\eqref{sk-median-theory}] is expected to be
valid for small $q$ only. The theory expression can be multiplied by an
(unknown) constant which has been
set to unity. All panels have the same horizontal and vertical scales.
}
\label{fig:int-mean-median}
\end{figure*}

There are two possible interpretations of the trends discussed above.
If one believes that the RSB picture holds for all of the models studied
here, then one can point to the fact that all the $\Delta$ curves are
nondecreasing and assert that they will all approach unity as $N
\rightarrow \infty$, just extremely slowly for the 3D EA model and the
1D $\sigma=0.896$ model.  An argument supporting this idea is made in
Ref.~\onlinecite{billoire:13} and rebutted in
Ref.~\onlinecite{yucesoy:13b}.  If on the other hand, one believes the
droplet scenario or the chaotic pair scenario holds for finite-dimensional
spin glasses, then the flattening of the curves for these models is a
prelude to an eventual decrease to zero.  Unfortunately, the sizes
currently accessible to Monte Carlo simulation do not permit one to
sharply distinguish between these competing hypotheses.  Using an exact
algorithm for the two-dimensional (2D) Ising spin glass with bimodal
disorder, Middleton\cite{middleton:13} shows that the crossover to
decreasing behavior for $\Delta(q_0,\kappa)$ in 2D does occur at large
length scales. He also shows within a simplified droplet model, that the
large length scales are needed to see the predictions of the droplet
scenario manifest in the 3D EA model.  Overall, we see that we need
larger sizes to unambiguously determine from $\Delta(q_0,\kappa)$
whether the droplet or RSB picture applies to 3D-like models.

\subsection{Median $I^\text{med}(q)$ and mean $I^\text{av}(q)$
cumulative overlap distribution}
\label{sec:median}

\begin{figure}
\center
\includegraphics[width=\columnwidth]{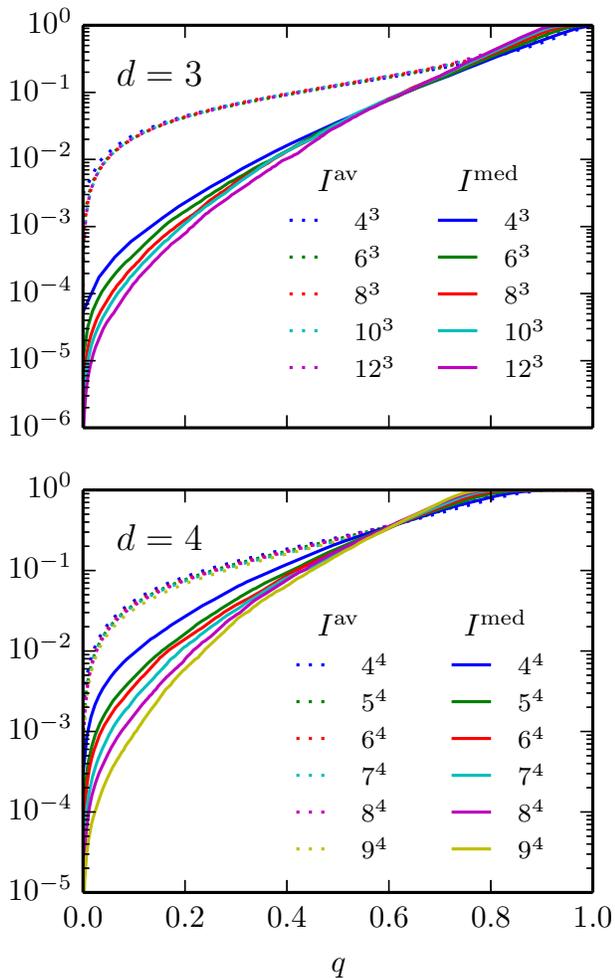}
\caption{(Color online)
Log-linear plot of $I^\text{med}(q)$ and $I^\text{av}(q)$ vs $q$ plot
for the 3D EA model at $T \simeq 0.42$ (upper panel) and for the 4D EA
model at  $T \simeq 0.90$ (lower panel).  Both panels have the {\em
same} horizontal scale.
}
\label{fig:mean-median-EA}
\end{figure}

In this section, we compare the mean $I^\text{av}(q)$ and the median
$I^\text{med}(q)$ of the cumulative overlap distribution.  Figure
\ref{fig:int-mean-median} shows results for $I^\text{av}(q)$ and
$I^\text{med}(q)$ for the SK model and several long-range models for
a temperature close to $0.4 T_c$.  Figure \ref{fig:mean-median-EA} shows
the same quantities for the 3D EA and 4D EA models.

As noted in earlier work, the results for the average show very little
size dependence for all models. This is a prediction of the RSB picture
which certainly applies to the SK model.  By contrast, in the droplet
picture $I^\text{av}(q)$ is predicted to vanish as\cite{fisher:86}
$L^{-\theta}$.  The observed independence of $I^\text{av}(q)$ with
respect to $L$ is one of the strongest arguments in favor of the RSB
picture for finite-dimensional Ising spin-glass models.  However, it has
been argued, e.g., Refs.~\onlinecite{moore:98,middleton:13}, that there
are strong finite-size corrections and that the asymptotic behavior
predicted by the droplet model for $I^\text{av}(q)$ would only be seen
for sizes larger than those accessible in simulations. This is why the
median has been proposed\cite{middleton:13} as an alternative to the
mean.

The data for the median of the SK model in
Fig.~\ref{fig:int-mean-median} show a rapid decrease at small $q$, which
is \textit{very strongly size dependent}.  As discussed in
Sec.~\ref{sec:qoi} above, the rapid decrease is expected in the RSB
picture since it predicts that $I^\text{med}(q)$ is exponentially small
in $1/q$ [see Eq.~\eqref{sk-median-theory}].  The theoretical result is
shown as a solid line in the SK panel.  It is plausible that the data
will approach the theory in the large $N$ limit, but there are strong
finite-size effects at small $q$ for the sizes that can be simulated, so
the data for the largest sizes are still far from the theoretical
prediction. This already indicates that the median is not a very useful
measure to distinguish the RSB picture from the droplet picture.

The median data for the long-range 1D model with $\sigma = 0.6$, which
is in the mean-field region, shows similar trends to that for the SK
model. On the other hand, for the long-range model furthest from mean-field 
theory, $\sigma = 0.896$, the data also decrease rapidly at small
$q$ but are less dependent on size. The data for the 3D and 4D EA models
in Fig.~\ref{fig:mean-median-EA} also show a rapid decrease at small $q$,
which is quite strongly size dependent.

We have seen that even for the SK model it would be very difficult to
extrapolate the numerical data to an infinite system size.  For the
long-range models, the most likely candidate for droplet theory
behavior, according to which the median (such as the average) vanishes in
the thermodynamic limit, is $\sigma = 0.896$. However, for this model,
the data are not zero for small $q$ and there is rather little size
dependence, implying that, if the droplet picture does hold, it will
only be seen for much larger sizes than can be simulated. This is the
same situation as for the mean (if the droplet picture is correct).
Consequently, it does not seem to us that the median of the cumulative
order parameter distribution is a particularly useful quantity to
distinguish the droplet and RSB pictures.

\subsection{Typical overlap distribution, $P^\text{typ}(q)$}
\label{sec:typical}

\begin{figure}
\center
\includegraphics[width=\columnwidth]{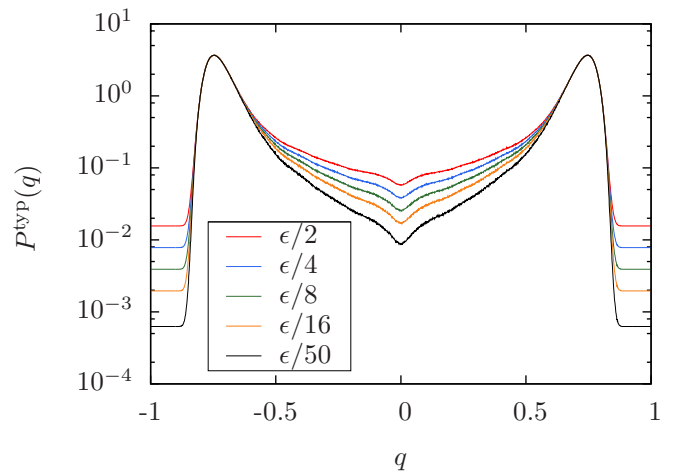}
\caption{(Color online)
Log-linear $P^{\rm typ}(q)$ plot for the SK model for $N=2048$ showing
the strong dependence on the zero-replacement value $\epsilon/k$.  See
the main text for details.
}
\label{fig:EP}
\end{figure}

Estimating $P^\text{typ}(q)$---defined in Eq.~\eqref{eq:typ} as the
exponential of the average of the logarithm---from Monte Carlo
simulations is problematic because the finite number of observations
means that the result can be precisely zero if the average is comparable
to, or smaller than, $\epsilon$, the inverse of the number of
measurements. Such results make the typical value undefined according to
Eq.\ \eqref{eq:typ}.  One can regularize this problem by replacing zero
values of $P_{\mathcal J}(q)$ with the small value $\epsilon/k$ for a
reasonable range of $k$, in the hope that the result would not be too
sensitive to the choice of $k$.  Unfortunately, there is a strong
dependence on $k$, as seen in Fig.~\ref{fig:EP}, where $P^\text{typ}(q)$
is plotted for several values of $k$ for the SK model for $N=2048$. The
dependence on $k$ indicates that $P^\text{typ}(q)$ cannot be reliably
measured in Monte Carlo simulations with feasible run lengths.

\section{Summary and Conclusions}
\label{sec:concl}

We have studied the overlap distribution for several Ising spin-glass
models using recently proposed observables.  We consider 1D long-range
models, 3D and 4D short-range (Edwards-Anderson) models, and the 
infinite-range (Sherrington-Kirkpatrick) model.  The three 
observables are
all obtained from the single-sample overlap distribution
$P_{\mathcal{J}}(q)$.  They are the fraction of peaked samples
$\Delta(q_0,\kappa)$, the integrated median $I^\text{med}(q)$, and the
typical value $P^\text{typ}(q)$.  These observables were proposed to
help distinguish between the replica symmetry breaking picture and
two-state pictures such as the droplet model.   While none of these
statistics unambiguously differentiates between these competing pictures,
it appears that $\Delta$ does the best job.  In particular,  there is a
qualitative distinction between the behavior for the 3D EA model and the
long-range 1D model with $\sigma=0.896$ that is expected to mimic it, on
the one hand, and the mean-field SK model and the 1D model with
$\sigma=0.6$ that is expected to be in the mean-field regime, on the
other hand.  For a reasonable range of $q_0$ and $\kappa$, the two
3D-like models do not show an increase in $\Delta$ for the largest sizes
while the mean-field models are sharply increasing for the largest
sizes.  The increase in $\Delta$ for the mean-field model is exactly
what we expect from the RSB picture.  The results for the 3D-like models
are ambiguous because eventually $\Delta$ must go either to zero or one.
It is possible that for much larger sizes $\Delta$ will begin to
increase, indicating RSB behavior, but simulating such large system
sizes at very low temperatures is unfeasible at present.

The other proposed measures do not appear to be useful in numerical
simulations for distinguishing scenarios.  The typical value of the
overlap $P^\text{typ}(q)$ cannot be measured in feasible Monte Carlo
simulations while the median value of the cumulative overlap
$I^\text{med}(q)$ is very small at small $q$ even for the SK model and
has a very strong size dependence. For the droplet model $I^\text{med}(q)$
is presumably zero at small $q$ for $N \to\infty$. However, the strong
size dependence of the results in this region of small $q$ makes it
impossible to tell numerically, if the data are going to zero or just to
a very small value, even for the SK model.  Curiously, there is
\textit{less} size dependence for the 3D model and the equivalent
1D with $\sigma = 0.896$ than for the SK model.

Recently, we became aware of a related paper by Billoire {\em et
al}.\cite{billoire:14a}.  Reference \onlinecite{billoire:14a} argues
that the data for $I^\text{med}(q)$ for the SK model ``converge nicely
to some limiting curve when $N$ increases'' and that ``trading the
average for the median does make the analysis more clear cut.'' In
contrast, we find a strong finite-size dependence for $I^\text{med}(q)$
for the SK model in the important small-$q$ region (clearly visible in a
logarithmic scale) and largely because of this we do not find that the
median is particularly helpful in distinguishing between the droplet and
RSB pictures.

\acknowledgments

We would like to thank A.~A.~Middleton, C.~Newman and D.~L.~Stein for
discussions.  M.W.~and A.P.Y.~acknowledge support from the NSF (Grant
No.~DMR-1207036). H.G.K.~acknowledges support from the NSF (Grant
No.~DMR-1151387) and would like to thank Rochelt for inspiration.
J.M.~and B.Y.~acknowledge support from NSF (Grant No.~DMR-1208046). We
thank the Texas Advanced Computing Center (TACC) at The University of
Texas at Austin for providing HPC resources (Stampede Cluster), ETH
Zurich for CPU time on the Brutus cluster, and Texas A\&M University for
access to their Eos cluster.

\bibliography{refs,comments}

\begin{thebibliography}{40}
\bibitem[{com({\natexlab{a}})}]{comment:norm}
\bibinfo{note}{Note that if $z \widetilde{p}_{i j} \ll 1$, then $p_{i j}$ in
  Eq.~\eqref{P_dil} is given by $p_{i j} = z \widetilde{p}_{i j}$, but
  otherwise there are corrections due to the rejection of pairs $i$, $j$ when
  there is already a bond between them.}
\end{thebibliography}

\end{document}